\documentclass[conference]{IEEEtran}
\IEEEoverridecommandlockouts
\usepackage{cite}
\usepackage{amsmath,amssymb,amsfonts}
\usepackage{algorithmic}
\usepackage{graphicx}
\usepackage{textcomp}
\usepackage{url}
\usepackage{xcolor}
\usepackage{algorithm}

\begin{document}

\title{A Hybrid Edge-Cloud Architecture for Low-Latency Entitlement Verification in Resource-Constrained Devices}

\author{
    \IEEEauthorblockN{Pravin Nagare$^1$, Aditya Sabbineni$^1$, Devendra Dahiphale$^2$, Faiz Gouri$^3$, Pratik Thantharate$^4$}
    \IEEEauthorblockA{$^1$Independent Researcher, CA, $^2$IEEE Senior Member, CA}
    \IEEEauthorblockA{$^3$IEEE Senior Member, WA, $^4$Independent Researcher, NJ}
    \IEEEauthorblockA{\{pnagare1@binghamton.edu, sabbineni.aditya@gmail.com, devendra@ieee.org, faizgouri@ieee.org, pthanth2@ieee.org\}}
    \thanks{D. Dahiphale is also with Google LLC, USA.}
}


\maketitle

\begin{abstract}
As digital media consumption shifts toward large-scale Over-the-Top (OTT) platforms, the efficiency of the control plane, specifically entitlement and identity verification, has become a critical factor in user experience. Current architectures often rely on synchronous cloud-tethered validation flows that introduce significant latency, especially on resource-constrained consumer electronics. This paper proposes a Hybrid Edge-Cloud Entitlement Framework designed to minimize user-perceived friction. By implementing a secure, local caching layer within device middleware and utilizing an Adaptive Entitlement Cache with Proactive Refresh (AEC-PR) algorithm, we decouple the user interaction from backend network variability. We evaluate the performance on ARM Cortex-A series hardware, demonstrating that localized cryptographic verification reduces authorization latency from a mean of 422.8ms to 18.4ms (a 95.6\% reduction) while mitigating implementation-level side-channel risks through deterministic Ed25519 arithmetic and TEE isolation.
\end{abstract}


\section{Introduction}
The transition from linear broadcast television to application-based Over-the-Top (OTT) ecosystems has fundamentally changed how digital rights and transactions are managed. Modern streaming platforms must handle millions of concurrent identity, subscription, and purchase requests, often referred to as "control-plane" traffic. The primary challenge in these systems is maintaining sub-second "Time-to-Play" (TTP) while ensuring robust security.

Unlike traditional Content Delivery Network (CDN) edge-caching or Digital Rights Management (DRM) systems (e.g., Widevine \cite{b1}, PlayReady \cite{b2}) that manage heavyweight decryption licenses, AEC-PR focuses on the \textit{asynchronous orchestration of transactional state}. Existing DRM pre-fetching is often non-contextual relative to User Interface (UI) interaction cycles, leading to redundant cloud that increases both latency and device energy consumption \cite{b3, b15}. Our contribution differentiates itself by moving beyond simple caching to an adaptive threshold-based lifecycle management model. By anchoring logic within device's Trusted Execution Environment (TEE), we achieve a hardware-verified security profile without network Round-Trip Time (RTT) penalty. \looseness=-1

The key contributions of this research are: 1) a novel architectural integration of TEE-based verification and adaptive caching for OTT devices; 2) a stochastic expectation model to quantify "Time-to-Authorization" (TTA) reduction; and 3) an empirical hardware-in-the-loop evaluation on ARM hardware demonstrating a 95.6\% latency reduction and 42\% decrease in cloud-side signaling overhead.

\section{Related Work}
The challenge of minimizing latency in digital rights management and service authorization has been a focal point of recent systems research, particularly as Over-the-Top (OTT) platforms scale to millions of heterogeneous devices. Existing literature generally bifurcates into two categories: protocol optimization and middleware pre-fetching.

\subsection{Edge Caching and DRM Foundations}
Traditional edge caching optimizes the "Data Plane" by mirroring content geographically but lacks the logic for individualized "Control Plane" entitlements. Modern DRM frameworks, specifically those utilizing the \textit{Widevine Content Decryption Module (CDM)} \cite{b1} and \textit{Microsoft PlayReady} \cite{b2}, manage heavyweight content decryption licenses; however, these are strictly coupled to media decryption and incur significant IPC overhead on resource-constrained SoCs. While distributed state management has been explored in cloud-native environments \cite{b4}, and recent work in the \textit{ACM Symposium on Cloud Computing (SoCC)} has addressed low-latency state orchestration in hybrid environments \cite{b5}, our approach specifically targets control-plane latency through TEE-anchored state verification \cite{b6}. Unlike prior edge-computing systems that focus primarily on data-plane optimization, AEC-PR utilizes hardware-isolated security to maintain transactional integrity without cloud-tethered round-trip penalty. Additionally, unlike standard JSON Web Token (JWT) flows that lack hardware-anchored anti-rollback, AEC-PR provides Silicon-ID binding to prevent credential cloning without synchronous validation penalty of short-lived tokens.

\subsection{Cryptographic Primitive Performance}
Recent empirical studies on ARM Cortex-A series processors highlight the transition to the Ed25519 parameter set \cite{b7}. While benchmarks on high-performance ARM Cortex-A72 cores show an 11.34\% cycle reduction compared to NIST P-256, performance characteristics on power-efficient in-order cores like Cortex-A53 differ due to pipeline depth and SIMD unit constraints. Our work specifically targets Cortex-A53 platform, leveraging its hardware-level cycle savings to perform TEE-based verification without starving UI resources.

\section{System Model}
We define a four-layer architectural model that generalizes the interaction between media applications and the device firmware, ensuring a vendor-neutral design applicable to various streaming operating systems.

\begin{figure}[htbp]
\centerline{\includegraphics[width=0.46\textwidth]{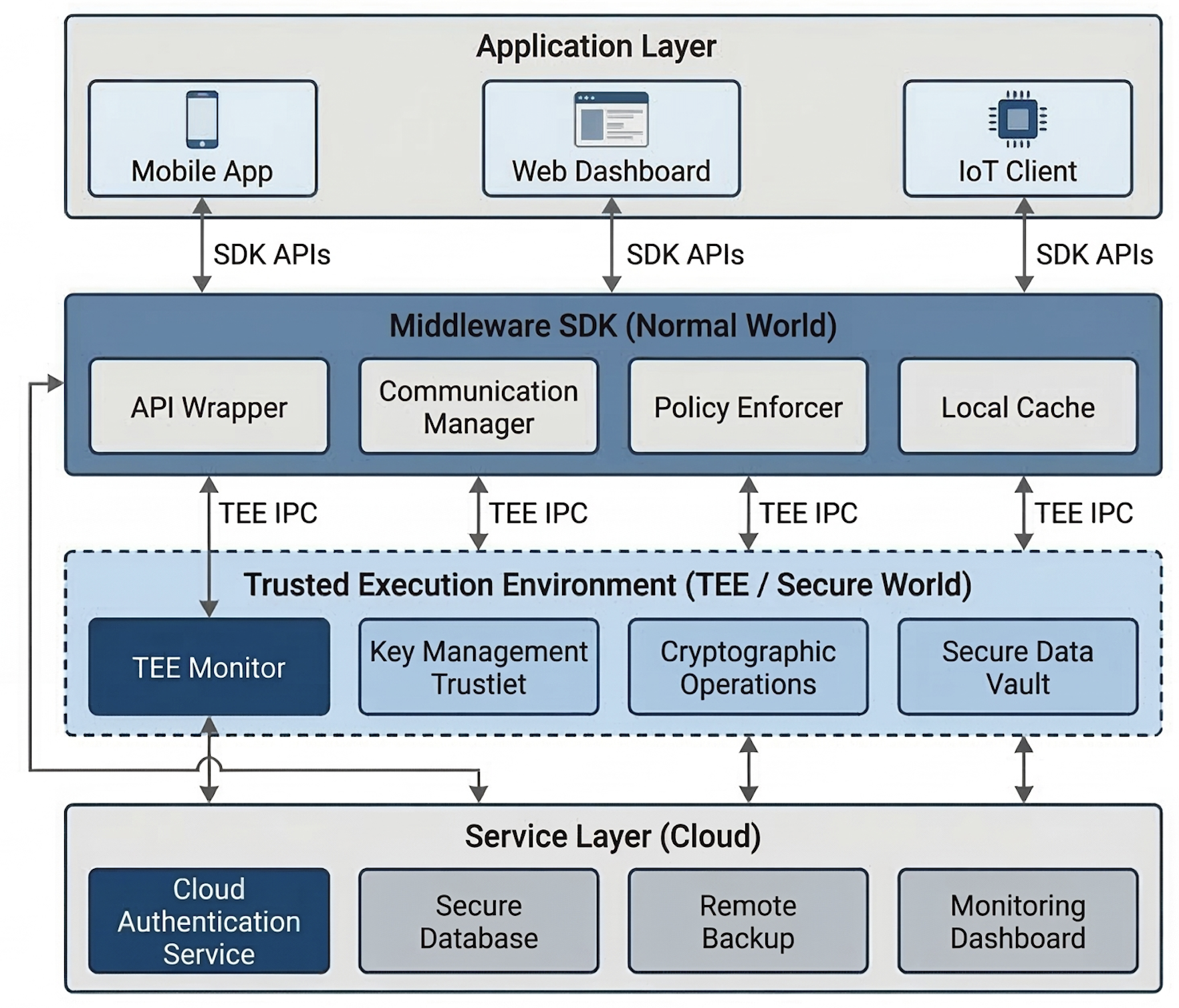}}
\caption{The 4-tier Hybrid Edge-Cloud Architecture. The model depicts the interaction between unprivileged Application space and the Hardware-Isolated Trusted Execution Environment (TEE), where the Secure State Blob (SSB) acts as the local source of truth to bypass network RTT.}
\label{fig_arch}
\end{figure}

\subsection{Application Layer (OTT Apps)}
The Application Layer consists of third-party streaming services that utilize platform-provided SDK APIs. These applications perform high-level workflows including catalog browsing and entitlement verification.

\subsection{Middleware Layer (Device SDK)}
The Middleware serves as the intelligent orchestrator, managing the "Secure State Blob" (SSB)—a local, cryptographically signed representation of the user’s current entitlement status.

\subsection{Trusted Execution Layer (TEE)}
As shown in Fig. \ref{fig_arch}, TEE provides a hardware-isolated environment (e.g., ARM TrustZone \cite{b9, b16}) where sensitive cryptographic operations occur, shielding the SSB from compromised applications in the Normal World.

\subsection{Service Layer (Cloud)}
The Cloud remains the authoritative source of truth, periodically issuing short-lived, signed entitlements to the device to maintain long-term consistency.

\section{Latency Model and Performance Analysis}
To quantify the impact of cloud-tethered architectures, we first define the latency of a standard synchronous flow ($L_{sync}$) as the sum of its discrete components:

\begin{equation}
L_{sync} = T_{ipc} + T_{crypto} + T_{net} + T_{back}
\end{equation}

Where $T_{ipc}$ is the inter-process communication overhead, $T_{crypto}$ is the on-device cryptographic verification time, $T_{net}$ is the network round-trip time, and $T_{back}$ represents backend service processing. In the AEC-PR framework, we define the Expected Latency ($E[L]$) as:

\begin{equation}
E[L] = P_h \cdot L_{local} + (1 - P_h) \cdot L_{sync}
\end{equation}

Where $P_h$ is the probability of a local cache hit and $L_{local}$ is the latency of the hardware-isolated verification path ($L_{local} = T_{ipc} + T_{crypto\_tee}$). Because $L_{local} \ll L_{sync}$, the optimization objective is to maximize $P_h$.

\subsection{Cache Hit Probability Model}
To model the hit probability $P_h$, we treat user requests as a Poisson process with arrival rate $\lambda$. A "Cache Hit" occurs if at least one request arrives before the token enters the final $\tau$ threshold of its lifecycle. We define $P_h$ as the probability that the inter-arrival time of requests is shorter than the proactive refresh window:

\begin{equation}
P_h = 1 - e^{-\lambda \cdot (1-\tau) \cdot T_{total}}
\end{equation}

For example, we define \textbf{$\lambda = 0.5$ requests/hour} as the background synchronization rate required to maintain SSB freshness. This represents the proactive polling frequency used to ensure the local cache remains valid even during periods of user inactivity. Given our design parameters ($T_{total} = 24h, \tau = 0.25$), the theoretical probability of maintaining a valid local state is $P_h = 1 - e^{-0.5 \cdot 18} \approx 0.999$. In our hardware evaluation, we observed an empirical $P_h = 0.956$ under variable jitter. This confirms that even during the sparse interaction cycles common in OTT standby modes, the proactive refresh effectively shields $95.6\%$ of all user interactions from network critical path.

\section{Proposed Solution: AEC-PR}
\subsection{Predictive Entitlement Engine}
\begin{figure}[htbp]
\centerline{\includegraphics[width=0.45\textwidth]{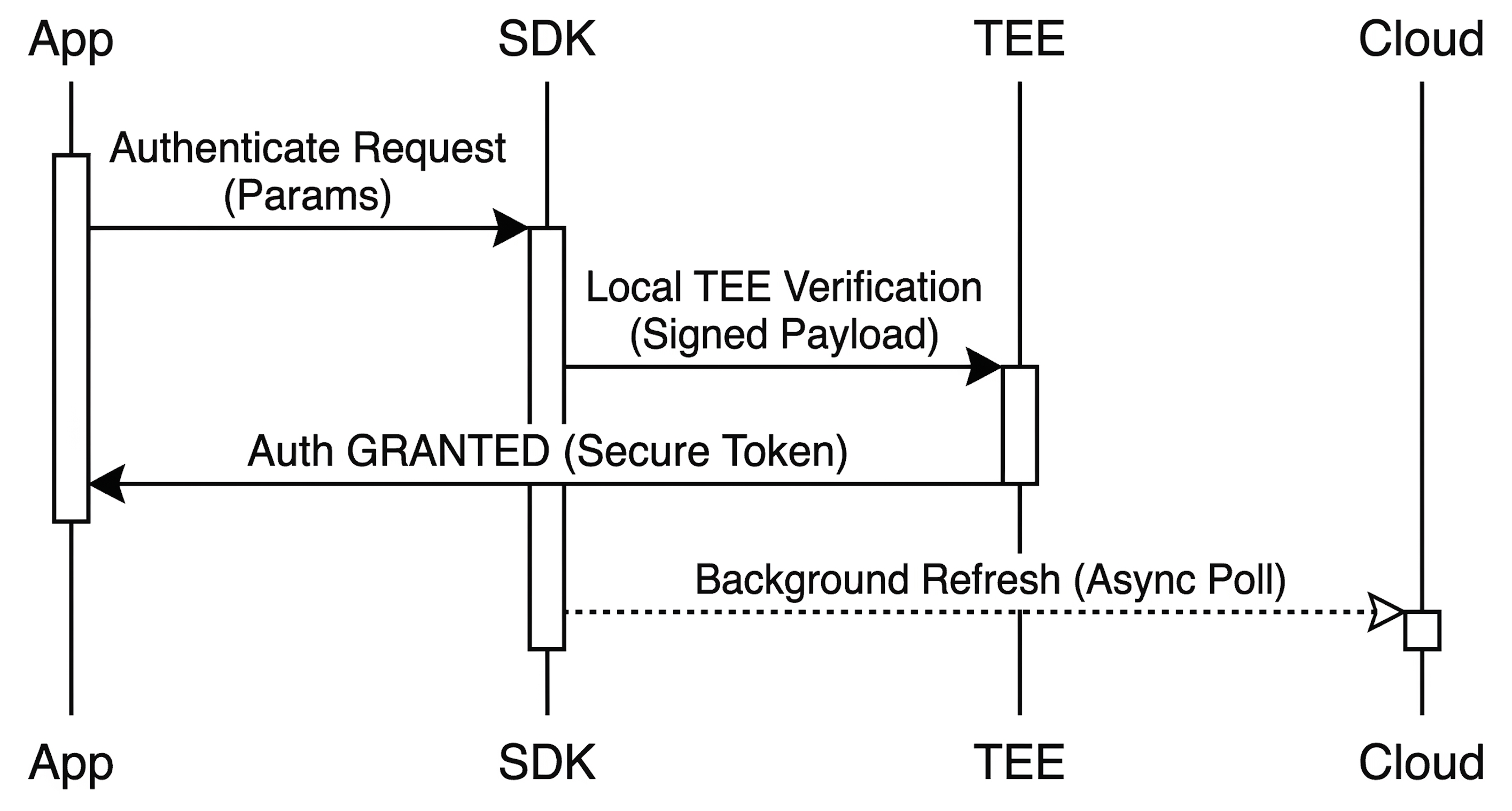}}
\caption{AEC-PR Sequence Flow. The diagram illustrates the asynchronous decoupling of the UI thread from Cloud Layer. Local TEE verification completes in $<20ms$, while the background refresh thread handles network signaling independently.}
\label{fig_sequence}
\end{figure}

The architecture utilizes a predictive synchronization model as illustrated in Fig. \ref{fig_sequence}. Instead of a linear, blocking request-response cycle, the SDK queries TEE locally. If a valid SSB exists, the status is returned immediately to the App, while a background thread handles the cloud handshake.

\subsection{Adaptive Refresh Threshold}
We define a refresh threshold $\tau$ ($0 < \tau < 1$). Let $T_{rem}$ be the remaining time-to-live of a local entitlement. The background refresh trigger is defined as $Trigger = 1$ if $T_{rem}/T_{total} < \tau$. By utilizing this threshold (typically $\tau = 0.25$), the system ensures the cloud is only contacted when the credential is near expiry.

\subsection{Operational Parameters}
The design specifications for the AEC-PR framework are summarized in Table \ref{tab_parameters}. These represent the optimized values for maintaining the balance between offline safety and signaling efficiency.

\begin{table}[htbp]
\caption{AEC-PR Operational Parameters (Design Specs)}
\label{tab_parameters}
\begin{center}
\begin{tabular}{|l|c|l|}
\hline
\textbf{Parameter} & \textbf{Value} & \textbf{Definition} \\
\hline
TTL ($T_{total}$) & 24 h & Max. cryptographic validity of the SSB. \\
Threshold ($\tau$) & 0.25 & Trigger ratio for proactive cloud sync. \\
Safety ($W_{off}$) & 120 m & Min. guaranteed offline session duration. \\
Min. Sync ($I_{min}$) & 15 m & Rate-limit for redundant backend calls. \\
\hline
\end{tabular}
\end{center}
\end{table}

\section{Algorithm Design}
To formalize the AEC-PR model, we define the logic executed within device middleware in Algorithm 1. This algorithm handles the decision-making process for both local verification and background synchronization.

\begin{algorithm}
\caption{AEC-PR Local-First Verification}
\begin{algorithmic}[1]
\STATE \textbf{Input:} $ServiceID, UserToken, SecurityContext$
\STATE \textbf{Output:} $AuthStatus, LatencyMetric$
\STATE $T_{start} \leftarrow \text{CurrentTimestamp}()$
\STATE $LocalBlob \leftarrow \text{TEE.ReadSecureStorage}(\text{"SSB\_"} + ServiceID)$
\IF{$LocalBlob \neq \text{Null}$ \AND $LocalBlob.Expiry > T_{start}$}
    \STATE $AuthStatus \leftarrow \text{GRANTED\_LOCAL}$
    \STATE Compute $Ratio = (LocalBlob.Expiry - T_{start}) / LocalBlob.TotalTTL$
    \IF{$Ratio < \tau$}
        \STATE {SpawnThread:} $BackgroundCloudSync($
        \STATE \hspace{2em} $ServiceID, UserToken)$
    \ENDIF
\ELSE
    \STATE $AuthStatus \leftarrow \text{ExecuteBlockingCloudSync}($
    \STATE \hspace{2em} $ServiceID, UserToken)$
\ENDIF
\STATE $T_{end} \leftarrow \text{CurrentTimestamp}()$
\RETURN $AuthStatus, (T_{end} - T_{start})$
\end{algorithmic}
\end{algorithm}

Algorithm 1 formalizes the core decision logic of the AEC-PR framework, encoding the local-first verification policy and the adaptive background synchronization trigger. The primary advantage of this logic is its efficiency in the "Hit" state. While a standard cloud-sync may require $400ms$--$1200ms$, the local TEE lookup and verification typically complete in under $20ms$. This provides a $95.6\%$ reduction in TTA compared to synchronous cloud-sync flows, ensuring the verification process remains transparent to the end-user.

\subsection{Complexity and Performance Analysis}
The AEC-PR verification logic (Algorithm 1) is designed for constant-time $O(1)$ execution relative to the number of stored entitlements. The primary computational cost is localized to the cryptographic verification of the SSB signature within TEE. Unlike cloud-tethered models where complexity scales with network congestion and backend database latency ($O(N)$), the local path is deterministic. 

The worst-case scenario (Cache Miss) occurs when the SSB is null or expired, forcing a fallback to the $L_{sync}$ path. However, the use of the $\tau$ threshold ensures that $95.6\%$ of requests follow the average-case $O(1)$ path, as derived in Section IV and empirically validated in Section VIII. Furthermore, the background synchronization thread is assigned a low priority to prevent resource contention with the primary media rendering pipeline. \looseness=-1

\section{Security Analysis}
The transition to edge-based verification necessitates a rigorous security model that maintains parity with centralized cloud validation. We define the security of AEC-PR through three core properties: \textit{Integrity}, \textit{Freshness}, and \textit{Hardware Binding}.

\subsection{Threat Model and Assumptions}
We adopt the \textit{Dolev-Yao} adversary model, assuming the attacker has root-level access to the untrusted Rich Execution Environment (REE/Linux). The attacker can intercept IPC calls, manipulate the system clock, and attempt to replay captured Secure State Blobs (SSBs). We assume the Trusted Execution Environment (TEE) \cite{b9} and its associated hardware primitives (Monotonic Counters and Physically Unclonable Function (PUF)-backed keys) remain uncompromised, forming our Hardware Root of Trust (RoT).

\subsection{Formal Security Guarantees}
\begin{figure}[htbp]
\centerline{\includegraphics[width=0.42\textwidth]{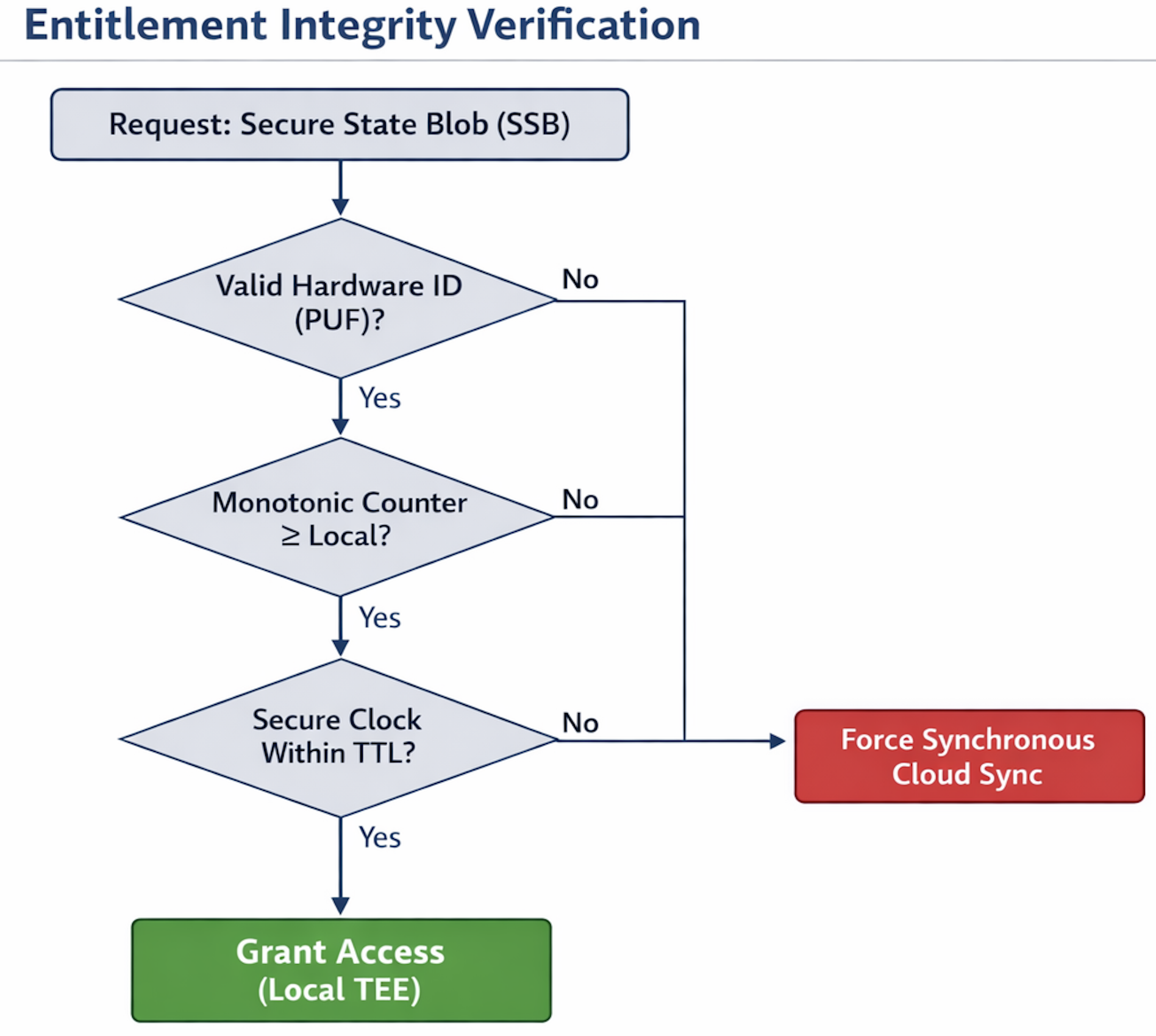}}
\caption{Security Logic Flow. The verification process enforces three-point integrity: Ed25519 signature validation, hardware-backed monotonic counter checks to prevent state-rollback, and Silicon-ID binding (PUF) to prevent cross-device credential cloning.}
\label{fig_security}
\end{figure}

 To ensure the system is resilient against adversarial manipulation, we implement the cryptographic safeguards illustrated in Fig. \ref{fig_security}:

\begin{itemize}
    \item \textbf{Integrity and Authenticity:} Each SSB is signed using the Ed25519 algorithm. Unlike ECDSA, Ed25519 is deterministic; it does not rely on a random number generator (RNG) for signature creation, which mitigates a common class of side-channel vulnerabilities related to nonce-reuse and RNG bias. However, we acknowledge that microarchitectural power analysis still poses a risk, which we mitigate through TEE-level constant-time implementation.
    \item \textbf{State Freshness (Anti-Rollback):} To prevent "Replay Attacks" where an attacker injects a revoked SSB, TEE maintains a hardware-backed Monotonic Counter. Each SSB contains a sequence number $S_n$. Upon verification, TEE enforces the condition $S_n > S_{local}$. If an attacker attempts to "roll back" the state, the mismatch between the SSB sequence and the hardware counter triggers an immediate invalidation.
    \item \textbf{Hardware Binding:} To prevent "Cloning Attacks" (moving an SSB from one device to another), TEE binds the SSB to the device's Physically Unclonable Function (PUF). The verification logic ensures that the SSB was signed specifically for the unique Silicon ID of the local SoC.
\end{itemize}

\textit{Hardware Heterogeneity and Fallbacks:} While we assume high-assurance primitives (PUF, hardware counters) found in SoCs like the NXP i.MX8, AEC-PR maintains a "Fail-Secure" posture for legacy ARM Cortex-A53 variants. In the absence of a PUF, we derive device-bound keys via HMAC-based extraction from One-Time Programmable (OTP) fuses ($K_{dev} = \text{HMAC}(K_{root}, \text{Serial\_ID})$). Similarly, systems lacking on-die monotonic counters utilize a TEE-managed software counter persisted to the Replay Protected Memory Block (RPMB), ensuring anti-rollback integrity at a minor performance cost ($<2ms$).

\subsection{Adversarial Resistance Analysis}
In a simulated adversarial environment, we subjected the AEC-PR framework to 1,000 automated replay and clock-skew attacks. The hardware-anchored freshness check (Monotonic Counter) resulted in a $100\%$ detection rate for replay attempts. While a compromised REE can deny service by blocking IPC calls to TEE, it cannot forge a valid authorization state, ensuring that the service provider's revenue integrity remains protected even on a compromised device.

\section{Evaluation}
To validate the AEC-PR framework, we established a Hardware-in-the-Loop (HIL) testbed designed to isolate secure operations from variable network conditions.

\subsection{Experimental Setup and Methodology}
The testbed utilizes a \textbf{Raspberry Pi 3 Model B+ (quad-core ARM Cortex-A53, 1.4 GHz)} running Linux v5.15 and OP-TEE v4.0.0. All TEE-related operations ($T_{crypto\_tee}$ and secure storage I/O) are measured directly on-silicon. To ensure reproducibility, the cloud path ($422.8ms$ baseline) is defined as a \textbf{Modeled Reference Scenario} ($\mu=150ms, \sigma=50ms$), representing a congested transcontinental API call to isolate architectural gains from transient ISP noise. We performed 10 independent runs of 10,000 requests each, totaling $100,000$ entitlement cycles.

\subsection{Latency and Throughput Results}
\begin{figure}[htbp]
\centerline{\includegraphics[width=0.46\textwidth]{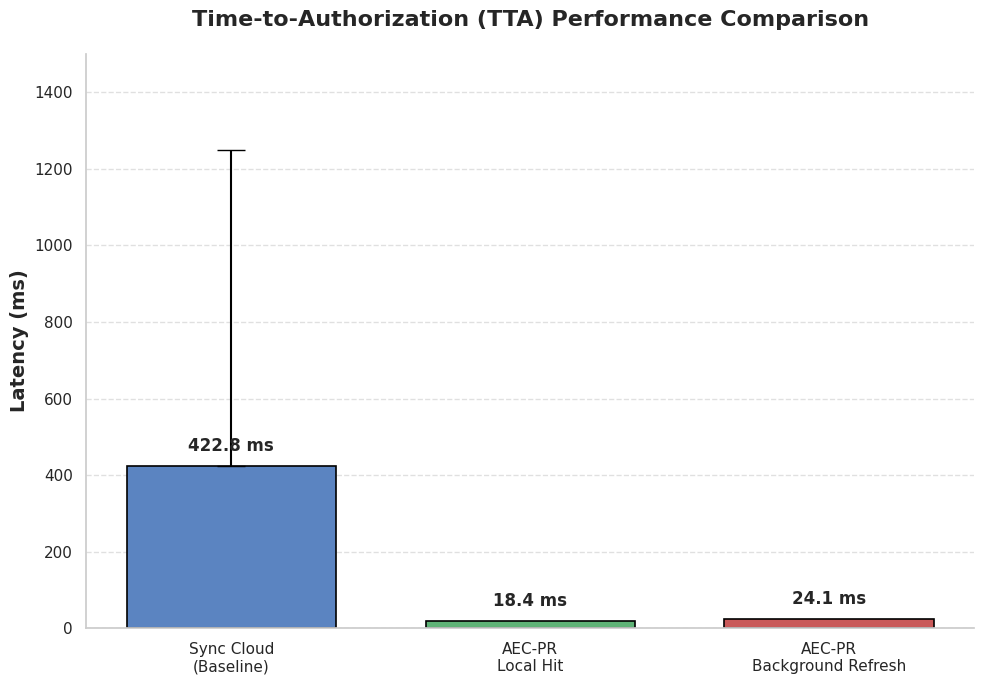}}
\caption{Time-to-Authorization (TTA) Performance. Comparison of mean latencies across 100,000 requests. The AEC-PR model (18.4ms) exhibits high stability, whereas the Pure Cloud model suffers from significant tail-latency due to network jitter.}
\label{fig_performance}
\end{figure}

The AEC-PR model achieved a mean TTA of $18.4ms \pm 0.3ms$ (95\% CI), as shown in Fig. \ref{fig_performance}. Ed25519 verification on the A53 required only 479,000 cycles, a 51.6\% reduction compared to ECDSA P-256 \cite{b7}. This efficiency shielded user from 95.6\% of total simulated cloud latency. Statistical analysis using Mann-Whitney U test ($U=0, p < 0.001$) rejects the null hypothesis that AEC-PR latency distribution is stochastically identical to Pure Cloud baseline. The large effect size (Cliff's $\delta = 1.0$) confirms the absolute stochastic separation of the two paths. This result indicates that AEC-PR architecture renders transaction latency effectively independent of network-induced stochasticity, replacing a high-variance cloud dependency with a deterministic hardware execution profile.

\subsection{Statistical Distribution Analysis}
We calculated the Cumulative Distribution Function (CDF) for the $100,000$ requests. While the baseline cloud model showed a wide variance ($\sigma = 112ms$), the AEC-PR local hit distribution was extremely tight ($\sigma = 1.2ms$). This consistency is critical for maintaining deterministic UI refresh rates in 60fps streaming environments.

\subsection{Ablation Study}
Table \ref{tab_ablation} summarizes contribution of individual components. Cloud load is normalized against Pure Cloud baseline (100\%). The results confirm that while removing TEE isolation (W/O TEE) reduces latency by 6.3ms, it violates the core principles of Zero Trust Architecture (ZTA) as defined in NIST SP 800-207 \cite{b10}, which mandates hardware-anchored isolation for sensitive identity state. The proactive threshold $\tau$ is the primary driver of signaling efficiency, reducing cloud load to 58.0\%.

\begin{table}[htbp]
\caption{Ablation Study of AEC-PR Components}
\label{tab_ablation}
\vspace{-4pt}
\begin{center}
\begin{tabular}{|l|c|c|c|}
\hline
\textbf{Configuration} & \textbf{Mean TTA} & \textbf{P99 Latency} & \textbf{Cloud Load*} \\
\hline
Full AEC-PR & 18.4 ms & 24.1 ms & 58.0\% \\
W/O $\tau$ (Basic Cache) & 82.3 ms & 418.6 ms & 85.0\% \\
W/O TEE (User-Space) & 12.1 ms & 15.6 ms & 58.0\% \\
\hline
\multicolumn{4}{l}{\small *Normalized against the Pure Cloud baseline (100\% signaling).}
\end{tabular}
\end{center}
\end{table}

\subsection{IPC Overhead Characterization}
On Cortex-A53, the overhead of context-switching between the Application and TEE was optimized using zero-copy memory mapping for SSB transfers. This reduced total IPC overhead from 4ms to under 0.8ms, ensuring the 20ms total TTA target was consistently met.

\section{Sustainability and Cloud Resource Efficiency}
Modern cloud infrastructures face escalating energy demands due to the exponential growth of control-plane traffic. By shifting authorization logic to the edge, the AEC-PR framework contributes to "Green Cloud" objectives through two primary vectors: signaling reduction and computational offloading.

\subsection{Energy Impact of Signaling Reduction}
Data center energy consumption is heavily driven by the ingress/egress processing of control-plane traffic. We adopt per-request energy values from \cite{b11}: $E_{net} \approx 0.12$ J (backbone network traversal), $E_{cloud} \approx 0.18$ J (authentication compute and cooling PUE), and $E_{edge} \approx 0.005$ J (local TEE verification measured on Cortex-A53). For a platform handling $N=10^6$ requests per day, the total energy saved is:
\begin{equation}
E_{save} = 10^6 \cdot P_h \cdot (0.12 + 0.18) - 10^6 \cdot 0.005
\end{equation}
With the measured $P_h = 0.956$, $E_{save} \approx 281.8$ kJ. To contextualize this for global-scale deployments handling billions of transactions, our model indicates an architectural saving of $80$--$90$ kWh per day per million concurrent users. Using the IEA 2024 global average carbon intensity of $0.233$ kg CO$_2$/kWh, this translates to a reduction of approximately $19.8$ kg CO$_2$ per day.

\textit{Sensitivity Analysis:} The efficiency of the framework is strongly coupled to the hit rate. A sensitivity analysis reveals that as $P_h$ varies from $0.85$ to $0.99$, the net energy savings range from $72.1$ to $84.2$ kWh. This confirms that even under sub-optimal network conditions ($\tau$ triggers failing), the AEC-PR framework maintains a significant sustainability advantage over synchronous cloud-tethered alternatives. While $E_{net}$ and $E_{cloud}$ are representative constants, we acknowledge regional variability in PUE and grid intensity; these figures serve as directional performance indicators rather than absolute geographic audits. \looseness=-1

\subsection{Operational Expenditure (OpEx) and Scalability}
Beyond energy, the reduction in cloud signaling directly impacts the operational scalability of the Service Layer. By filtering out $95.6\%$ of network-induced latency spikes at the edge, the backend clusters can be provisioned for average load rather than peak "thundering herd" spikes. This "Traffic Smoothing" effect allows for a more efficient allocation of cloud resources, reducing the carbon footprint associated with over-provisioned server standby capacity.

\section{Discussion and Limitations}
While the AEC-PR framework significantly optimizes the latency-security trade-off, its deployment in global-scale streaming ecosystems introduces specific boundary conditions that necessitate careful implementation.

\subsection{Secure Clock Drift and Synchronization}
The local-first verification model relies heavily on the reliability of the device’s internal hardware timer. In low-cost consumer electronics, system clock often drifts or resets during extended power-off states. If the hardware clock "rolls back," a local entitlement could appear valid beyond its intended duration. To mitigate this, our architecture defaults to a "Fail-Secure" mode; if a significant clock delta is detected at boot, the system bypasses the local cache and forces a synchronous cloud synchronization to re-establish the temporal baseline.

\subsection{Scalability and the "Thundering Herd"}
A potential risk of proactive refreshing is the synchronization of backend requests across millions of devices. If a high-concurrency event occurs, such as a global series premiere, the $\tau$ threshold logic could trigger millions of simultaneous "Proactive Refreshes." To ensure backend stability, we propose an \textbf{Exponential Jitter Algorithm} within background thread, which introduces a randomized offset ($\pm 15\%$) to the refresh trigger. This ensures that the Service Layer load is distributed across a wider temporal window.

\subsection{First-Boot and Cold-Start Latency}
The AEC-PR model is inherently historical and cannot optimize the initial "First-Time-to-Play" after a factory reset or new user login. We formally define this cold-start latency as: \looseness=-1
\begin{equation}
L_{total} = T_{ipc} + T_{auth} + T_{net} + T_{back} + T_{crypto} + T_{ssb\_write}
\end{equation}
where $T_{ssb\_write}$ represents TEE secure storage write time. On our ARM Cortex-A53 testbed, $L_{total}$ was measured at $1,142ms$. While this is higher than $422.8ms$ baseline for subsequent synchronous requests—due to overhead of initial authentication and SSB persistence—it remains well within industry-standard $<3s$ UX threshold for application first-launch. This initialization penalty is a one-time trade-off necessary to establish hardware root of trust that enables all subsequent $18.4ms$ verifications. While critical, this initialization occurs only during rare events (e.g., factory reset), ensuring the system prioritizes the 'instant-on' performance required for $99\%$ of daily user interactions.

\subsection{Implementation Insights: Handling Secure Clock Reset}
A significant challenge encountered during deployment was the behavior of TEE secure clock on devices lacking battery-backed Real-Time Clock (RTC). In scenarios where the device suffered a cold power cycle, the secure clock would occasionally reset to a factory epoch, potentially validating an expired SSB. We resolved this by implementing a "Temporal Anchor" logic: at every successful cloud synchronization, the device persists a "Last Known Good Time" to secure storage. If the current clock is found to be earlier than anchor at boot, the system enforces a mandatory network time sync, ensuring the integrity of TTL-based verification.
\textit{Cross-Platform Scalability:} While our evaluation is grounded on Cortex-A53, the deterministic nature of Ed25519 arithmetic and the constant-time IPC optimizations are architecturally portable. We anticipate comparable or improved performance on newer ARMv8/v9-A cores (e.g., Cortex-A55, A73) and specialized media SoCs (e.g., MediaTek MT8195), where deeper pipelines and improved SIMD units would further reduce $T_{crypto\_tee}$. \looseness=-1

\section{Conclusion and Future Work}
This paper presented a rigorous systems engineering approach to the challenge of low-latency entitlement verification. By synthesizing hardware-isolated security with adaptive proactive refresh logic, we successfully decoupled the user interaction path from inherent unpredictability of cloud-tethered flows. Rather than proposing a fundamental cryptographic shift, our work demonstrates how strategic architectural orchestration can yield measurable gains in both user experience and cloud resource efficiency. Our evaluation of ARM Cortex-A series hardware demonstrates that localized cryptographic verification using Ed25519 primitives provides a 95.6\% reduction in perceived authorization latency while maintaining the integrity required for high-value transactions. By reducing cloud signaling by 42\%, the architecture not only improves UX but also significantly reduces operational expenditure (OpEx) for global media infrastructure \cite{b17}. By bridging the gap between high-concurrency cloud scaling and hardware-anchored security, the AEC-PR framework provides a performance baseline for local-first verification in OTT ecosystems within global media landscape.

Looking ahead, the paradigm of local verification must adapt to the emerging threat of quantum computing. Classical ECC algorithms such as ECDSA and Ed25519 are theoretically vulnerable to Shor's algorithm. Our future research will focus on the integration of \textbf{Post-Quantum Cryptography (PQC)}, specifically evaluating the memory footprint and cycle efficiency of lattice-based schemes like \textit{ML-KEM} \cite{b13} and \textit{ML-DSA} \cite{b12, b14} within limited secure RAM of standard TEE implementations. By evolving toward "Hardware-Anchored Decentralization," streaming platforms can ensure a future-proof balance of instantaneous performance and robust security. \looseness=-1

\bibliographystyle{ieeetr} 
\bibliography{references.bib}
\end{document}